\documentclass[a4paper]{article}

\usepackage{INTERSPEECH2021}

\usepackage{amsmath,graphicx}
\usepackage{multirow}
\usepackage{setspace}
\usepackage{algorithm}
\usepackage{algorithmic}
\usepackage{epstopdf}
\usepackage{balance}

\title{Efficient Conformer with Prob-Sparse Attention Mechanism for End-to-End Speech Recognition}
\name{Xiong Wang$^{1*}$\thanks{$^{*}$The first two authors contributed equally to this work.},  Sining Sun$^{2*}$, Lei Xie$^{1}$\sthanks{~~Corresponding author.}~~, Long Ma$^{2}$}
\address{
	$^1$Audio, Speech and Language Processing Group,\\
	School of Computer Science, Northwestern Polytechnical University, Xi'an, China\\
	$^2$Tencent, Beijing, China}
\email{xwang@npu-aslp.org, siningsun@tencent.com, lxie@nwpu.edu.cn, malonema@tencent.com}

\begin{document}

\maketitle
\begin{abstract}
End-to-end models are favored in automatic speech recognition (ASR) because of their simplified system structure and superior performance. Among these models, Transformer and Conformer have achieved state-of-the-art recognition accuracy in which self-attention plays a vital role in capturing important global information. However,  the time and memory complexity of self-attention increases squarely with the length of the sentence. In this paper, a prob-sparse self-attention mechanism is introduced into Conformer to sparse the computing process of self-attention in order to accelerate inference speed and reduce space consumption. Specifically, we adopt a Kullback-Leibler divergence based sparsity measurement for each query to decide whether we compute the attention function on this query. By using the prob-sparse attention mechanism, we achieve impressively 8\% to 45\% inference speed-up and 15\% to 45\% memory usage reduction of the self-attention module of Conformer Transducer while maintaining the same level of error rate.

\end{abstract}
\linespread{0.3}
\section{Introduction}
In recent years, end-to-end (E2E) automatic speech recognition (ASR) technology has made great progress with its simplified architecture and competitive performance. Transducer~\cite{graves2012sequence,battenberg2017exploring,rnntshibie,graves2014towards,wang2021cascade,zhang2021tiny} and attention based encoder-decoder (AED)~\cite{transformer,chan2016listen,dong2018speech} are two popular E2E frameworks. For both kinds of models, the encoder  is crucial for good performance. The Transformer architecture with self-attention has recently become the core block of most E2E models. Compared with RNN with long-short term memory (LSTM) units, Transformer block gives better accuracy and efficient computation because of its ability of modeling long-range global context. However, vanilla self-attention based Transformer block is less capable of capturing local information, which is also essential to speech recognition. To take advantage of both local and global contexts, Conformer~\cite{conformer} block is proposed, which combines convolution with self-attention, where the convolution is used to capture the local information while the self-attention for the global information consideration. Much work has shown that Conformer can obtain better speech recognition accuracy with fewer model parameters. 

For both Transformer and Conformer blocks, self-attention plays a very important role. With the self-attention mechanism, each output is the weighted combination of the whole sequence, which makes it have the ability to model the global information. However, there are two problems to apply global self-attention to the ASR task. The first one is the computation efficiency. The time complexity of self-attention can reach $O (T^2)$ concerning the sequence length $T$. With the increase of the input sequence length, the time complexity will increase in a quadratic way. For the ASR task, the speech signal can last from seconds to minutes and the input length varies from dozens to thousands in terms of frames. For example, in the popular benchmark dataset — LibriSpeech~\cite{panayotov2015librispeech}, the audio clips last from about 1.5 to 20 seconds. Therefore, reducing computational complexity  is essential, especially for decoding long utterances. The second problem is that the self-attention mechanism does not consider the  information redundancy issue.  It is clear that considering the whole input can benefit the current output. However, the speech signal is highly structured and there can be redundant frames that are useless for a specified input. 

Many researchers have recently attempted to optimize the above mentioned computation and memory consumption of self-attention mechanism. For example, Longformer~\cite{beltagy2020longformer} was proposed to handle the long document processing tasks, which combines local windowed attention with a task motivated global attention to process documents of thousands of tokens or longer. As for the ASR task, Shi et al. recently proposed Emformer~\cite{shi2020emformer}, which converts a long-range history context into an augmentation memory bank to reduce the computation complexity of self-attention. Although these previous attempts can reduce the computation complexity, they have not considered the second issue $-$ computation redundancy.  


In this paper, inspired by the Informer~\cite{zhou2020informer} originally proposed for long sequence time-series forecasting, we propose an efficient Conformer with prob-sparse attention mechanism  to tackle the above mentioned two problems. Specifically, the attention score between $i$-th query and all keys can be defined as a probability. If attention score of the $i$-th query over all keys is close to a uniform distribution, the self-attention degenerates into a trivial sum of all values, which is apparently redundant. Therefore, only the queries with attention score distribution far from uniform distribution play dominant roles. Under this probability distribution measurement, prob-sparse self-attention only allows each key to attend to the top-$u$ queries. Compared with the vanilla self-attention mechanism, the query matrix is sparse because it only contains top-$u$ dominant queries. The prob-sparse method will reduce the computation complexity into $O(uT)$ without performance degradation because this process retains effective queries and removes redundant computation. Furthermore, to reduce the extra computation of the sparse measurement, we also exploit ``sparsity measurement sharing'' strategies between layers.  By using the prob-sparse attention mechanism with sparsity measurement sharing, we achieve impressively 10\% to 40\% inference speed-up of the Conformer Transducer while maintaining the same level of CER/WER on AISHELL-1~\cite{aishell1} and LibriSpeech~\cite{panayotov2015librispeech} corpora.

\section{Conformer Transducer}\label{sec2}
\subsection{Architecture of Transducer}
\begin{figure}
	\centering
	\includegraphics[width=0.8\linewidth]{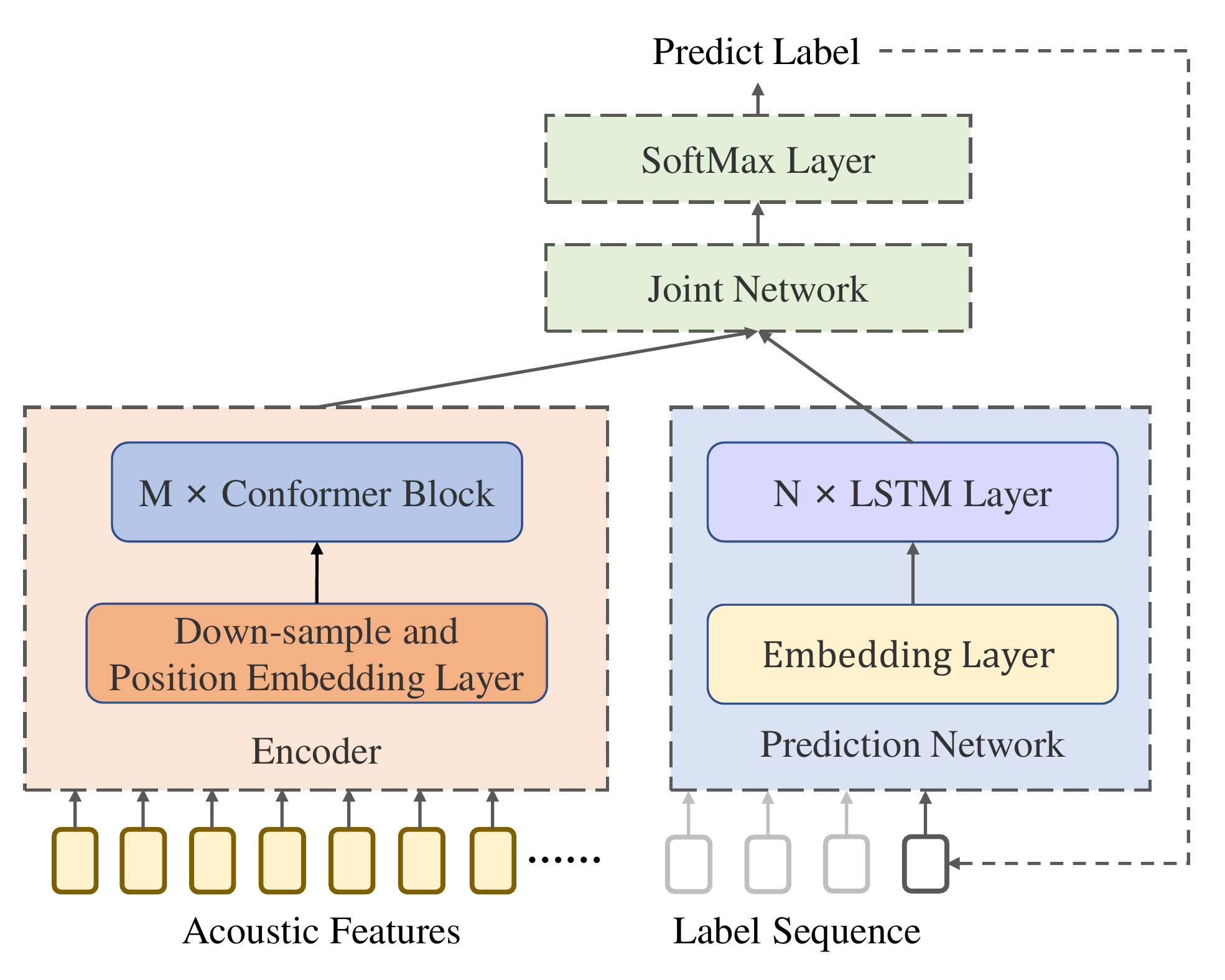}\vspace{0pt}
	\caption{Architecture of Conformer Transducer.}\vspace{-0pt}
	\label{fig:1}
	\vspace{-5mm}
\end{figure}

In this paper, we use Conformer Transducer to verify our proposed method. The Transducer model can establish the alignment between acoustic features $\mathbf{x}=[{{\mathbf{x}}_{1}},{{\mathbf{x}}_{2}},\cdots ,{{\mathbf{x}}_{T}}]$ and label sequences $\mathbf{y}=[{{y}_{1}},{{y}_{2}},\cdots ,{{y}_{U}}]$, where $T$ is the number of feature frames and $U$ is the length of the label sequence. Figure~\ref{fig:1} shows the transducer structure used in this paper, which mainly includes an encoder, a prediction network and a joint network. Specifically, the encoder converts the acoustic features into a high-dimensional representation ${\mathbf{h}}^{enc}$ which can be expressed as
\begin{equation}
	{{\mathbf{h}}^{enc}}=\text{Encoder}(\mathbf{x}).
\end{equation}
The main purpose of prediction network is to generate a higher-dimensional representation $h_{u}^{pred}$ of last label ${y}_{u-1}$ as shown in Eq.~(\ref{eq2}). To avoid the sparsity caused by directly using one-hot label as input, an embedding layer is arranged before $N$ layers of LSTM.
\begin{equation}\label{eq2}
	h_{u}^{pred}=\text{Prediction}({{y}_{u-1}})
\end{equation}
The joint network is represented by several fully connected layers, and finally a softmax layer is used to predict the probability $P(k|t,u)$ of the next label, as shown in
\begin{equation}
	h_{t,u}={{\mathbf{W}}^{joint}}\tanh (\mathbf{U}h_{t}^{enc}+\mathbf{V}h_{u}^{pred}+b)+{{b}_{joint}}
\end{equation}
\begin{equation}
	P(k|t,u)=\text{softmax}(h_{t,u})
\end{equation}
where ${\mathbf{U}}$ and ${\mathbf{V}}$ are the projection matrices to combine $h_{t}^{enc}$ and $h_{u}^{pred}$, and ${\mathbf{W}}^{joint}$ is used to project the output dimension to the number of labels. During training, transducer uses the forward-backward algorithm~\cite{graves2013speech} to maximize the posterior of ${\bf{y}}$ given $\bf{x}$.

\subsection{Conformer Encoder}

Recently, Conformer has achieved excellent performance in speech recognition~\cite{conformer}. Compared with Transformer, Conformer adds an additional convolution module to make up for the lack of local feature extraction ability of single self attention. A standard Conformer block consists of four modules: feed-forward module, self-attention module, convolution module and another feed-forward module. The Conformer block used in this paper does not use relative positional embedding in the self-attention block. The specific composition of each module is described in \cite{conformer}.

As shown in Figure~\ref{fig:1}, when using Conformer as the encoder of Transducer, the input feature $\mathbf{x}$ first passes through the down-sampling layer and position embedding layer and then passes through $M$ Conformer blocks. Each Conformer block needs to go through a macaron-like feed-forward module (FFN), multi-head self attention module (MHSA), convolution module (Conv), and another FFN with layernorm in order to get a higher-dimensional representation. This process can be described in Eq.~(\ref{eq5}).
\begin{equation}\label{eq5}
	\begin{aligned}
		{x}_{FFN_1} &=x+\frac{1}{2} \mathrm{FFN}\left(x\right) \\
		x_{MHSA} &={x_{FFN_1}}+\operatorname{MHSA}\left(x+\frac{1}{2} \mathrm{FFN}\left(x\right)\right) \\
		x_{Conv} &=x_{MHSA}+\operatorname{Conv}\left(x_{MHSA}\right) \\
		x_{FFN_2} &=\text { LayerNorm }\left(x_{Conv}+\frac{1}{2} \mathrm{FFN}\left(x_{Conv}\right)\right)
	\end{aligned}
\end{equation}
And for the prediction network, it still has $N$-layer LSTM commonly used by Transducer, while joint network is also implemented by the standard fully-connection layer.

\section{Prob-Sparse Self-Attention}
\subsection{Self-attention Mechanism and Sparsity Analysis}
 Given input matrix $\mathbf{X}\in\mathbb{R}^{L\times d_{x}}$, where $L$ and $d_x$ are sequence length and dimension of input respectively. The self-attention mechanism first projects $\mathbf{X}$ into query matrix ${\mathbf{Q}=\mathbf{X}\mathbf{W}_q}$ , key matrix ${\mathbf{K}=\mathbf{X}\mathbf{W}_k}$ and value matrix ${\mathbf{V}=\mathbf{X}\mathbf{W}_v}$, where $\{\mathbf{Q},\mathbf{K}, \mathbf{V} \}\in \mathbb{R}^{L\times d}$ and $\mathbf{W}_q, \mathbf{W}_k, \mathbf{W}_v$ are projection matrices with appropriate shapes. Then attention in scaled dot-product form can be represented by

\begin{equation}\label{eqattn}
	\mathcal{A}(\mathbf{Q}, \mathbf{K}, \mathbf{V})=\operatorname{softmax}\left(\frac{\mathbf{Q} \mathbf{K}^{\top}}{\sqrt{d}}\right) \mathbf{V}.
\end{equation}

In order to describe prob-sparse self-attention, Eq.~(\ref{eqattn}) will be reformulated as its vector form. Specifically, for the $i$-th query $\mathbf{q}_i$ from $\mathbf{Q}$, the attention score over the $j$-th key from $\mathbf{K}$ can be expressed as  

\begin{equation}\label{selfeqattn1}
	p(\mathbf{k}_j|\mathbf{q}_i) = \frac{\mathrm{e}^{\mathbf{q}_{i} \mathbf{k}_{j}^{\top}} / \sqrt{d}}{\sum_{l=1}^{L} \mathrm{e}^{\mathbf{q}_{i} \mathbf{k}_{l}^{\top}} /_{\sqrt{d}}}.
\end{equation}
Here we define $p(\mathbf{k}_j|\mathbf{q}_i)$ as attention score of query $\mathbf{q}_i$ over key $\mathbf{k}_j$. Then the self-attention of $\mathbf{q}_i$ over $\mathbf{K}$ can be expressed as 
\begin{equation}\label{selfeqattn2}
	\mathcal{SA}(\mathbf{q}_i, \mathbf{K}, \mathbf{V})=\sum_{j=1}^{L}  p(\mathbf{k}_j|\mathbf{q}_i) \mathbf{v}_{j}.
\end{equation}

In this process, the time complexity of attention mechanism is $O(L^2)$. In fact, there is a certain sparsity in the query matrix~\cite{zhou2020informer}, which means that it is unnecessary to calculate attention for all queries. Therefore, vanilla self-attention will have a lot of redundant calculations.   
Generally speaking, if the distribution of $p(\mathbf{k}_j|\mathbf{q}_i)$ yields to uniform a distribution, that is $p(\mathbf{k}_j|\mathbf{q}_i)=1/L$,  the self-attention of Eq.~(\ref{selfeqattn2}) shrinks into a trivial average of all values. Therefore, only the queries whose attention scores over all keys are far from uniform distribution are important. 
From this aspect, the Kullback-Leibler(K-L) divergence between the real distribution $P$ of  $p(\mathbf{k}_j|\mathbf{q}_i)$   and the uniform distribution $U$ can be used to  measure the validity of each query, expressed by 
\begin{equation}\label{kl}
KL(P||U)=\ln \sum_{j=1}^{L} e^{\frac{\mathbf{q}_{i} \mathbf{k}_{j}^{\top}}{\sqrt{d}}}-\frac{1}{L} \sum_{j=1}^{L} \frac{\mathbf{q}_{i} \mathbf{k}_{j}^{\top}}{\sqrt{d}} - lnL.
\end{equation}
After removing the constant, we can get Eq.~(\ref{sparse}) and define it as sparsity measurement of $\mathbf{q}_i$. 
The queries with larger $M_{sparse}$ values play more important roles in the self-attention mechanism. In this way, we can measure the overall sparsity of each query and ignore the queries with lower $M_{sparse}$ values during self-attention calculation, which can speed up the whole calculation procedure. 
\begin{equation}\label{sparse}
M_{sparse}\left(\mathbf{q}_{i}, \mathbf{K}\right)=\ln \sum_{j=1}^{L_{K}} e^{\frac{\mathbf{q}_{i} \mathbf{k}_{j}^{\top}}{\sqrt{d}}}-\frac{1}{L} \sum_{j=1}^{L_{K}} \frac{\mathbf{q}_{i} \mathbf{k}_{j}^{\top}}{\sqrt{d}}
\end{equation}
\subsection{Prob-Sparse based Self-Attention}
Although the sparsity measurement proposed in Eq.~(\ref{sparse}) works well in theory, it needs to traverse all queries to calculate the dot-product pairs, which is still computing intensive. According to~\cite{zhou2020informer}, to further reduce the computation of this part, Eq.~(\ref{sparse}) can be approximated by  Eq.~(\ref{saprse_simple_sample}) using sampling methods:
\begin{equation}\label{saprse_simple_sample}
{\bar{M}_{sparse}}\left(\mathbf{q}_{i}, \tilde{\mathbf{K}}\right)=\max _{j}\left\{\frac{\mathbf{q}_{i} \mathbf{k}_{j}^{\top}}{\sqrt{d}}\right\}-\frac{1}{\tilde{L}} \sum_{j=1}^{\tilde{L}} \frac{\mathbf{q}_{i} \mathbf{k}_{j}^{\top}}{\sqrt{d}}
\end{equation}
where $\tilde{\mathbf{K}}$ is a random sampling key matrix and $\tilde{L}$ is sampling number defined as $\tilde{L} = r_{sample}\operatorname{ln}L$. Here, sampling rate $r_{sample}$ is constantly used to control how many samples can be sampled. After obtaining  $\bar{M}_{sparse}$ for each query, only top-$L_{sparse}$ queries with high $\bar{M}_{sparse}$ value will be used to calculate self-attention, where $L_{sparse}=r_{sparse}L$ and $0<r_{sparse}<1$ where $r_{sparse}$ is sparse rate. Finally, the self-attention will be calculated using

\begin{equation}\label{sparse_dot}
\mathcal{SA}(\mathbf{q}_i, \mathbf{K}, \mathbf{V})=\left\{\begin{array}{cc}
\sum_{j=1}^{L}  p(\mathbf{k}_j|\mathbf{q}_i) \mathbf{v}_{j}, & \text { if } i \in I_{sparse} \\
\mathbf{v}_{i}, & \text { else }
\end{array}\right.
\end{equation}
where $I_{sparse}$ is the index of  top-$L_{sparse}$ queries.  
It is obvious that for multi-layer attention, a query whose index is not in $I_{sparse}$ will directly output the value and attention calculation is no longer needed.
\section{Experiments}\label{sec4}
\subsection{Datasets}
In this paper, we verify our proposed prob-sparse attention mechanism based Conformer Transducer on two open source datasets: AISHELL-1 and LibriSpeech. The AISHELL-1 corpus contains 178 hours of labeled Mandarin speech data, recorded on different devices, and internally divided into the training, development and test sets~\footnote{www.openslr.org/33/}. LibriSpeech contains 970 hours of labeled English speech data, internally divided into training, clean and other development and test sets~\footnote{www.openslr.org/12/}.
\subsection{Experimental Setup}
For all experiments, we use 80-dimensional log mel-filterbanks plus the 3-dimensional pitch features extracted by the KALDI toolkit~\cite{povey2011kaldi} as the acoustic features . We also apply SpecAugment~\cite{park2019specaugment} for acoustic data augmentation. A set of 4230 Chinese characters is used as the modeling units for AISHELL-1 and a set of 5000 word pieces is used for LibriSpeech. All models are trained using the ESPNet toolkit~\cite{watanabe2018espnet}. During training, we use the Noam optimizer with peak learning rate at 1e-3 and gradient clipping is set at 5.0 for all models. All the models for evaluation are obtained by averaging the best five models according to the loss on the development set. In different hyper-parametric experiments, the random seed and training steps are the same. All the experimental results are conducted on 4 NVIDIA Tesla V100 GPUs.

The Conformer encoder consists of 2 convolutional layers to do 4-time down-sampling with position embedding layer and 16 Conformer blocks. The attention dimension of each Conformer block is 256 with 4 attention heads. The kernel size of convolution module is 3 and the dimension of feed-forward module is 1024. The prediction network has 1 LSTM layer with 256 hidden units and the joint network has 512 hidden units. The size of embedding layer is 256. For the  prob-sparse self-attention, it has two additional super parameters, one is the sampling rate $r_{sample}$ used to calculate the sparsity measurement, and the other is the sparse rate $r_{sparse}$ used to do scaled dot-product with sparse query. The influence of these two hyper-parameters will be analyzed in the later experiment.

\subsection{Training Strategy and Performance Analysis}
Table~\ref{tab:1} shows the results of our proposed prob-sparse self-attention based Conformer Transducer on AISHELL-1 test sets. B0 is our baseline model trained with vanilla self-attention from scratch, which means we compute the attention function on all the queries. E0 is trained with our proposed prob-sparse self-attention from scratch. Compared with baseline B0, E0 with prob-sparse self-attention obtains worse CER. The main reason behind the accuracy degradation is that prob-sparse mechanism depends on a relative reliable model to calculate $M_{sparse}$ for each query. Therefore, in experiment E1, we initialize our model from B0 and retrain the model using prob-sparse self-attention mechanism with sparse rate $r_{sparse}=0.5$, which means that we only compute the attention function on a half of queries for each Conformer block. Compared with B0, E1 obtains even better recognition performance with CER lowered from 6.7 to 6.5, which also proves that the information redundancy issue  we mentioned before does exist. To further exclude that the performance improvement comes from more training steps, experiment B1 is conducted, which proves that more training steps can not give further improvement. Furthermore, in order to show the effectiveness of our prob-sparse self-attention method, we also implement experiment E2 by sampling queries with random style. We can see that the CER increases from 6.5 (E1) to 8.2 (E2), which means that choosing queries randomly will hurt the model performance because it can not select the most useful queries. 

\begin{table}
\caption{ CER comparison on AISHELL-1 test sets with different parameter initialization styles and query selection ways. $r_{sample}=5$ for experiment with prob-sparse self-attention.}
	\label{tab:1}
	\centering
	\scalebox{0.9}{
\begin{tabular}{ccccc}
\hline
ExpID  & \begin{tabular}[c]{@{}c@{}}Param.\\ Initialization\end{tabular} & \begin{tabular}[c]{@{}c@{}}Selection \\ of $\mathbf{q}_i$\end{tabular} & $r_{sparse}$ & CER(\%) \\ \hline \hline
B0               & Random                                                 & All                  & 1.0                                                  & 6.7     \\ \hline
B1               & B0                                                     & All              & 1.0                                                      & 6.7     \\ \hline
E0               & Random                                              & Prob-sparse            & 0.5                                                   & 7.2     \\ \hline
E1              & B0                                                     & Prob-sparse         & 0.5                                                    & \textbf{6.5}     \\ \hline
E2               & B0                                                     & Random selection             & 0.5                                           & 8.2     \\ \hline
\end{tabular}
}
\vspace{-7mm}
\end{table}

\begin{figure}[!h]
	\centering
	\includegraphics[width=0.8\linewidth]{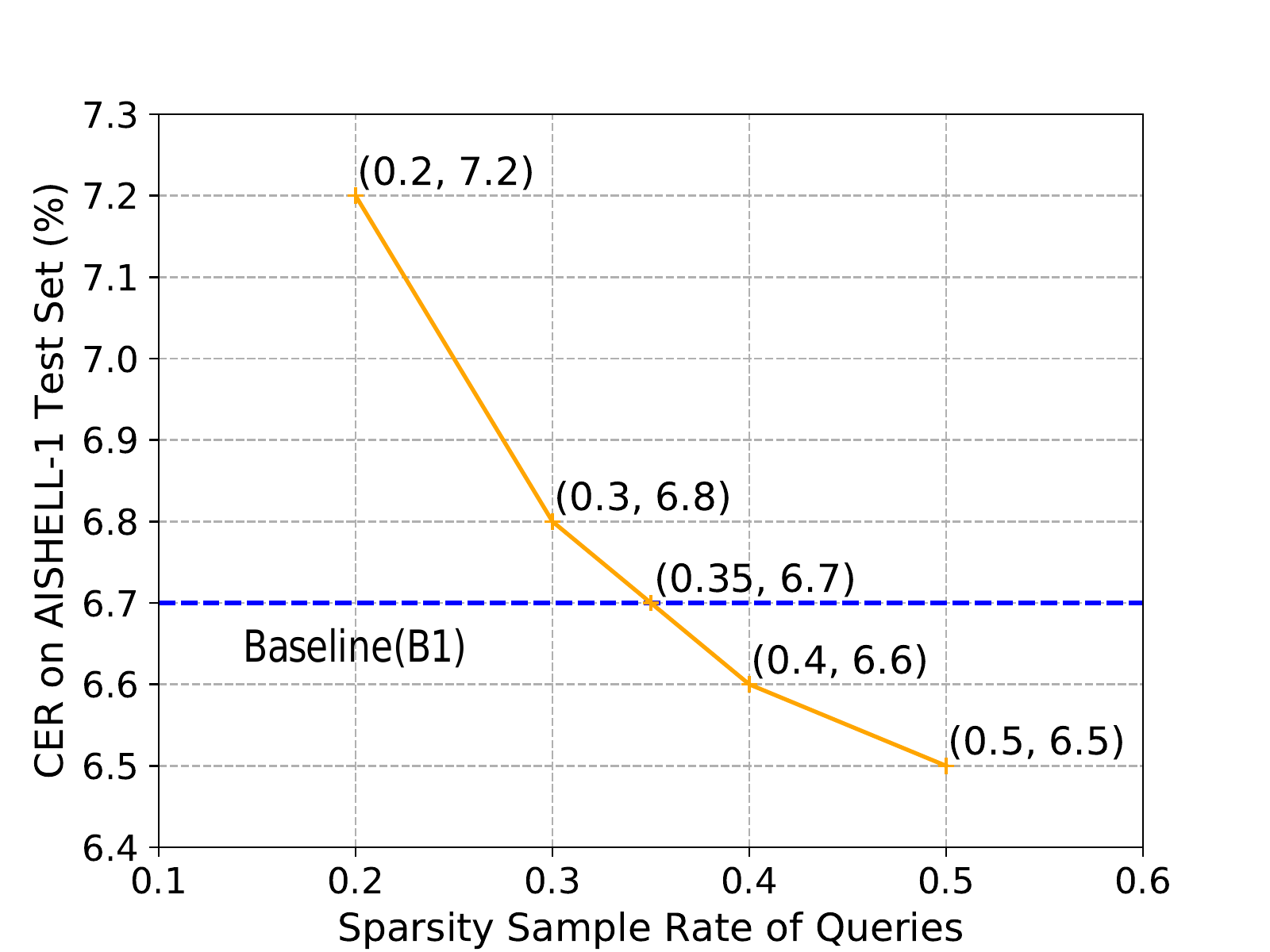}\vspace{-2mm}
	\caption{Relationship between CER and $r_{sparse}$ values on AISHELL-1 test set.}
	\label{fig:2}
	\vspace{-1mm}
\end{figure}

All the experiments in Table~\ref{tab:1} use sparse rate $r_{sparse}=0.5$. As analyzed in Section 3, small sparse rate means we compute the attention function on fewer queries. However, a small sparse rate also means more information will be discharged. To explore how sparse rate influence the model performance, we further conducted experiments with  several $r_{sparse}$ values. Figure~\ref{fig:2} shows how CER varies with respect to  $r_{sparse}$.  As we expected, when $r_{sparse}<0.35$, the CER is worse than our baseline model, which means too much information is ignored. On the contrary, when  $r_{sparse}$ is between 0.35 to 0.5, the prob-sparse attention method can obtain consistent better results than baseline because the model can discard redundant information efficiently. 
\vspace{-4mm}
\vspace{-5mm}
\subsection{Sparsity Measurement Share Through Layers}
As we mentioned before, in order to calculate the sparsity measurement $M_{sparse}$, some extra computations in each layer are introduced as described in Eq.~(\ref{saprse_simple_sample}). It's worth noticing that the computation complexity will be further reduced if the sparsity measurement $M_{sparse}$ can be shared through layers. To verify this hypothesis, we conduct experiments on AISHELL-1. Table~\ref{tab:3} shows results of $M_{sparse}$ sharing through layers. $N_{share}$ means that $M_{sparse}$ is calculated every $N_{share}$ layers and it will be shared for the next $N_{share}-1$ layers. In order to further reduce the calculation, we also try to reduce the sample number of keys in Eq.~(\ref{saprse_simple_sample}), which is controlled by $r_{sample}$. In Table~\ref{tab:3}, E1 is same as the one in Table~\ref{tab:1}, where $M_{sparse}$ will be calculated in each layer. For S0, we first reduce the queries sample rate factor from 5 to 1, which brings a small CER increase~(from 6.5 to 6.7). However,  we can  still obtain the same results as our baseline model B0 in Table~\ref{tab:1} even when $r_{sample}=1$. When we apply measurement share with $N_{share}=4$ for this 16-layer Conformer, we can reduce 4 times of calculation with no CER increasment, which verifies that the sparsity measurement can be shared through layers. Even when we increase  $N_{share}$ to 8 in S2, there is only negligible CER increase. Therefore, sparsity measurement share is an efficient way to reduce extra calculation for our ASR model.  


\begin{table}[!htb]
	\vspace{-0pt}
	\caption{Comparison of the results on AISHELL-1 test sets in CER with different ${r_{sample}}$ and ${N_{share}}$ values with ${r_{sparse}=0.5}$.}
	\vspace{0pt}
	\label{tab:3}
	\centering
	\scalebox{0.9}{
		\begin{tabular}{cccc}
			\hline 
			Exp ID & ${r_{sample}}$ & ${N_{share}}$ & CER(\%) \\ 
			\hline 
			\hline
			E1 & 5 & 1 & 6.5 \\ 
			\hline 
			S0 & 1 & 1 & 6.7 \\ 
			\hline 
			S1 & 1 & 4 & 6.7 \\ 
			\hline 
			S2 & 1 & 8 & 6.8 \\ 
			\hline 
		\end{tabular} 
	}
	
\end{table}

\subsection{Experiments on LibriSpeech}
 We also carry out experiments on LibriSpeech with different hyper parameters to verify our proposed method. LibriSpeech English dataset has a larger amount of data than AISHELL-1, which can well verify the reliability of the experiments we have done above. The experimental results are shown in Table~\ref{tab:4}. Note that compared with the results reported in~\cite{conformer}, our baseline results of M0 in Table~\ref{tab:4} are worse because we did not spend too much effort on model refinement. Based on the results in Table~\ref{tab:4},  we can achieve consistent conclusions with those on AISHELL-1. 
\begin{table}[!htb]
	\vspace{-0pt}
	\caption{Experimental results on LibriSpeech. L0-L3 are initialized by M0.}
	\vspace{0pt}
	\label{tab:4}
	\centering
	\scalebox{0.75}{
		\begin{tabular}{cccccccc}
			\hline
			\multirow{2}{*}{Exp ID} & \multirow{2}{*}{$r_{sample}$} &  \multirow{2}{*}{$r_{sparse}$} & \multirow{2}{*}{${N_{share}}$} & dev & dev & test & test \\
			  & & & & clean & other & clean & other \\
			\hline
			\hline
			M0 & / & 1 & /   & 4.0 & 10.3 & 4.2 & 10.3 \\
			\hline
			L0 & 5 & 0.5 & 4 & 4.0 & 10.5 & 4.3 & 10.5 \\
			\hline
			L1 & 5 & 0.5 & 8 & 4.1 & 10.8 & 4.4 & 10.9 \\
			\hline
			L2 & 1 & 0.5 & 4 & 4.1 & 10.6 & 4.3 & 10.7 \\
			\hline
			L3 & 1 & 0.5 & 8 & 4.2 & 10.9 & 4.4 & 11.0 \\
			\hline
		\end{tabular}}
\end{table}
\vspace{-5mm}
\subsection{Efficiency and Complexity}
\begin{figure}[!h]
	\centering
	\vspace{-10pt}
	\includegraphics[width=0.8\linewidth]{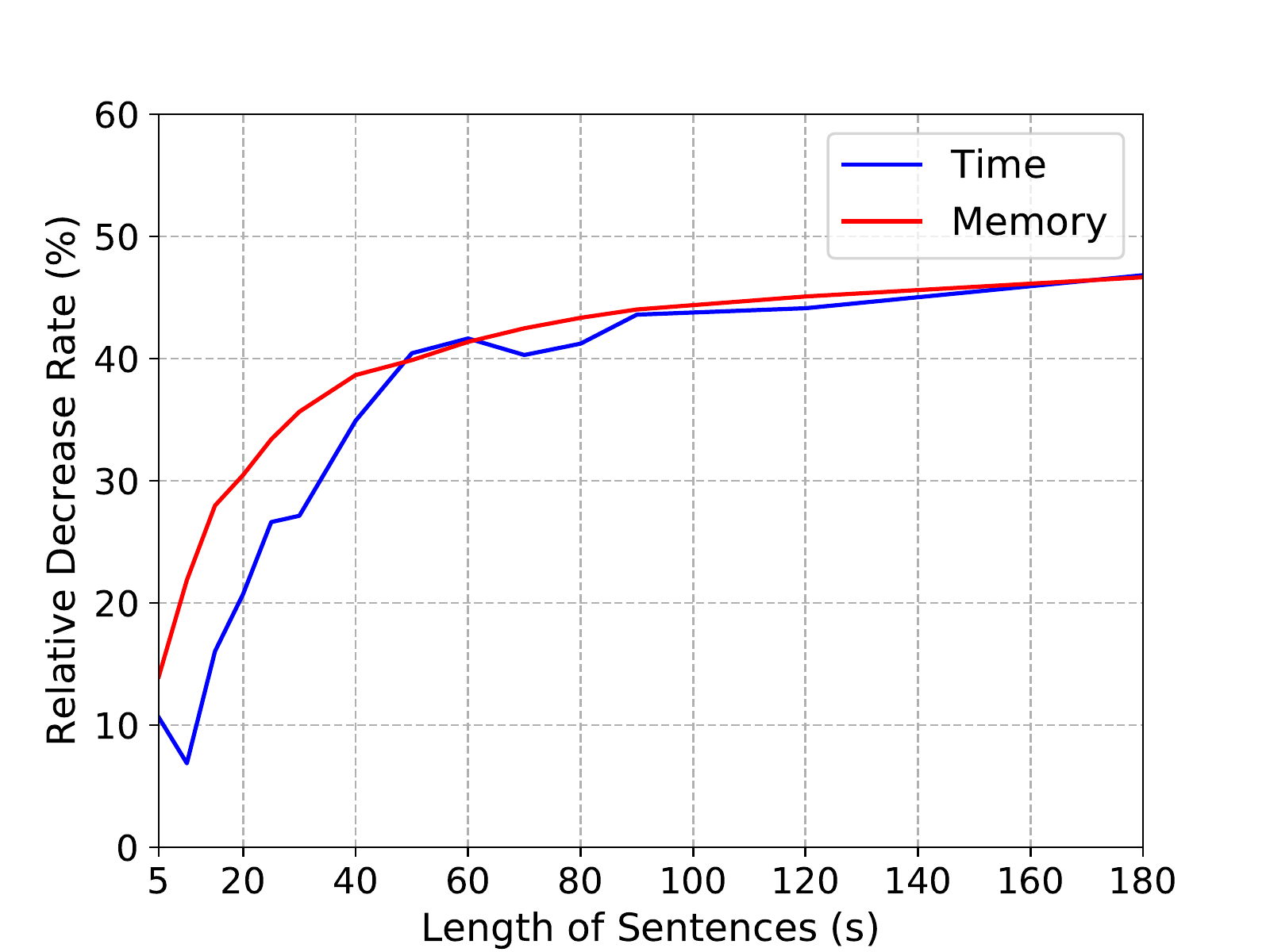}
	\caption{Time and memory usage between standard and prob-sparse self-attention module of Conformer-Transducer on L2 in Table~\ref{fig:3}. All results are evaluated on a single thread of Intel(R) Xeon(R) CPU E5-2680 v4 @ 2.40GHz. }\vspace{-0pt}
	\label{fig:3}
	\vspace{-6mm}
\end{figure}

From the above experiments, we verify that there is a certain sparsity of queries in the attention mechanism. After we use sparsity measurement to keep only the effective queries for calculation, we observe that the model performance will not degrade. At the same time, this operation can also greatly reduce the time and space complexity of self-attention. Specifically, we evaluate the time and memory consumption of the self-attention module for sentences with different length in the Conformer Transducer model during the decoding process. 
As shown in Figure~\ref{fig:3}, our proposed method can relatively increase the inference speed by 8\% to 45\% and reduce the memory usage by 15\% to 45\% in different sentence lengths. This advantage will be more obvious with the increase of sentence length.
\section{Conclusions}
In this paper, we proposed an efficient Conformer with prob-sparse attention mechanism. Specially, K-L divergence is adopted as a sparsity measurement to evaluate the effectiveness of each query in the self-attention module of Conformer and only more effective queries contribute to the result. We achieve impressively 8\% to 45\% inference speed-up and 15\% to 45\% memory usage reduction of the self-attention module of Conformer Transducer while maintaining the same level of error rate.

\bibliographystyle{IEEEtran}

\bibliography{template}


\end{document}